\documentclass{jfm}
\usepackage{graphicx}
\usepackage{color}
\usepackage{epstopdf, epsfig}
\usepackage[intlimits]{amsmath}

\shorttitle{Surfactant-Laden Bursting Bubbles}
\shortauthor{C. R. Constante-Amores et al.}

\title{Dynamics of a surfactant-laden bubble bursting through an interface}

\author{C. R. Constante-Amores\aff{1},
  L. Kahouadji\aff{1}\corresp{\email{l.kahouadji@imperial.ac.uk}},
  A. Batchvarov\aff{1}, 
  S. Shin\aff{2}, \ns J. Chergui\aff{3}, \ns D. Juric\aff{3} \and O. K. Matar\aff{1}}
  
\affiliation{
\aff{1}Department of Chemical Engineering, Imperial College London, London SW7 2AZ, UK 
\aff{2}Department of Mechanical and System Design Engineering, Hongik University, Seoul 121-791, Republic of Korea
\aff{3}Laboratoire d'Informatique pour la M\'ecanique et les Sciences de l'Ing\'enieur (LIMSI), Centre National de la Recherche Scientifique (CNRS), Universit\'e Paris Saclay, B\^at. 507, Rue du Belv\'ed\`ere, Campus Universitaire, 91405 Orsay, France
}

\begin{document}

\maketitle

\begin{abstract}
We study the effect of surfactant on the dynamics of a bubble bursting through an interface. We perform fully three-dimensional direct numerical simulations using a hybrid interface-tracking/level-set method accounting for surfactant-induced Marangoni stresses, sorption kinetics, and diffusive effects. We  select an initial bubble shape corresponding to a large Laplace number and a vanishingly small Bond number in order to neglect gravity, and isolate the effects of surfactant on the flow. Our results demonstrate that the presence of surfactant affects the dynamics of the system through Marangoni-induced flow, driving motion from high to low concentration regions, which is responsible for the onset of a recirculation zone close to the free surface. These Marangoni stresses  rigidify the interface, delay the cavity collapse, and influence the jet breakup process.
\end{abstract}

\section{Introduction}
When a bubble is resting close to a liquid-gas interface, its rupture gives rise to the formation of a central jet. 
This jet breaks up into small droplets, which could transport biological material, toxins, salts, surfactants or dissolved gases \citep{Woodcock_nature_1953,MacIntyre_jgr_1972,Veron_arfm_2015,Zenit_pt_2018,Seon_epjst_2017,Poulain_prl_2018}.
It is unsurprising therefore that the bursting bubble phenomena have received significant interest due to their occurrence in a multitude of natural and industrial applications. In the absence of contaminants, the physical mechanisms of surfactant-free bursting bubbles on the ejection of droplets have been widely studied experimentally \citep{Woodcock_nature_1953,Ghabache_pof_2014,Ghabache_prf_2016,Seon_epjst_2017}, numerically \citep{Deike_prf_2018,Gordillo_jfm_2019,Singh_prf_2019}, and through theoretical approaches \citep{Zeff_nature_2000,Ganan-Calvo_prl_2017,Lai_prl_2018,Blanco-Rodriguez_jfm_2020}.
 
This previous work has shown that when a nucleated bubble rises through the liquid and then rests close to a free surface, its static resting shape  results from a balance between gravitational and surface tension forces. The bubble shape may be characterised  by a submerged interface, a liquid layer/cap above the bubble, and an interface, which extends away from the bubble cap \citep{Toba_josj_1959,Ghabache_PhD_2015}. The cap is characterised by a length scale $\delta / R_o \sim O (10^{-6})$, where $\delta $ and $R_o$ refer to the liquid layer thickness and the initial bubble radius, respectively. The layer curvature creates over-pressure relative to the liquid bulk below it. The fluid within the layer drains over time enabling van der Waals forces to rupture  the interface when $\delta \rightarrow 0$ forming a hole \citep{Lhuissier_jfm_2012}. The hole leaves an open, unstable cavity that  collapses to form a vertical jet whose dynamics are governed by inertial, viscous, and capillary forces. 

Surfactants can affect the dynamics of surface-tension driven flows by the reduction of the local surface tension and capillary pressure, and by the formation of Marangoni stresses brought about by gradients in interfacial surfactant concentration. Studies involving surfactant effects on interfacial flows have received considerable attention, however, to the best of our knowledge, the influence of surfactants on the dynamics of  bursting bubbles has not been explored, and this is the subject of the present article. Here, we use three-dimensional numerical simulations to account for surfactant solubility, diffusion, sorption kinetics, and Marangoni stresses on the bursting phenomena. The rest of this paper is structured as follows: in Section \ref{Numerical}, the numerical method, governing dimensionless parameters, problem configuration, and validation, are introduced. Section \ref{sec:Results} presents the results, and concluding remarks are given in Section \ref{sec:Con}.

\section{Problem formulation and numerical method\label{Numerical}}
%
\begin{figure}
\begin{center} 
\includegraphics[width=0.95\linewidth]{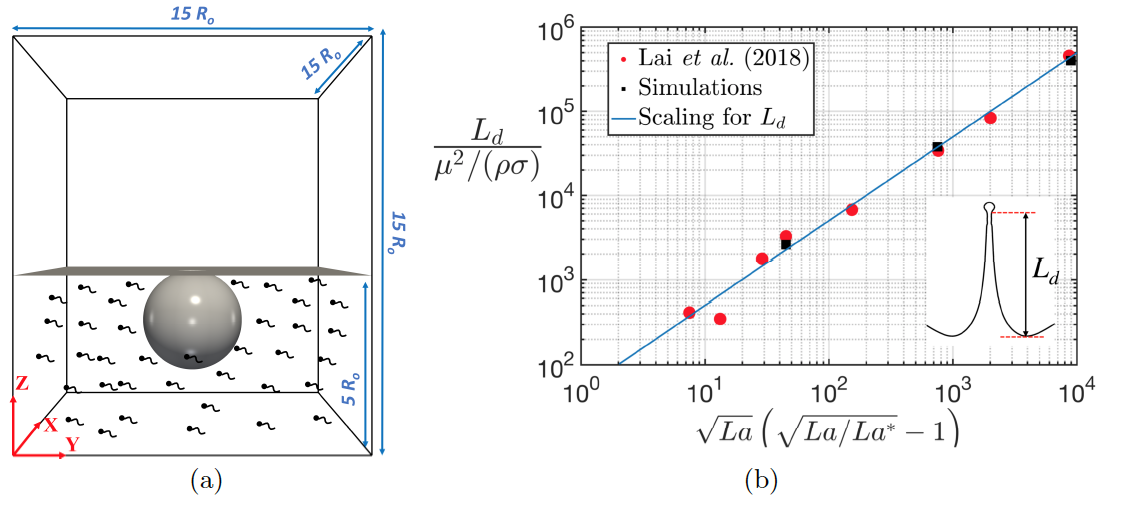}
\end{center} 
\caption{(a) Initial shape of the bubble resting close to the interface, highlighting the computational domain of size $(15 R_o)^3$ (not-to-scale) in a three-dimensional Cartesian domain $\mathbf{x} = (x, y, z)$, where Cartesian resolution is $768^3$; (b) comparison of surfactant-free simulations from the current study (squares) with scaling argument of the length of the jet (solid line) and numerical simulations (dots) proposed by \cite{Lai_prl_2018}.\label{configuration}} 
\end{figure}

The numerical simulations were performed by solving the two-phase Navier-Stokes equations in a three-dimensional Cartesian domain $\mathbf{x} = \left(x, y, z \right)$ (see figure \ref{configuration}). A hybrid front-tracking/level-set method was used to treat the interface, where surfactant transport was resolved both in the bulk and on the interface 
\citep{Shin_jcp_2018}. Here, and in what follows, all variables will be made dimensionless (represented by tildes) using
\begin{equation}
\quad \tilde{\mathbf{x}}=\frac{\mathbf{x}}{R_o},
\quad \tilde{t}=\frac{t}{T}, 
\quad \tilde{\textbf{u}}=\frac{\textbf{u}} {U},
\quad \tilde{p}=\frac{p}{\rho U^2}, 
\quad \tilde{\sigma}=\frac{\sigma}{\sigma_s},
\quad \tilde{\Gamma}=\frac{\Gamma}{\Gamma_\infty},
\quad \tilde{C}=\frac{C}{C_{\infty} },
\quad \tilde{C}=\frac{C_s}{C_{\infty} },
\end{equation}

\noindent	
where, $t$, $\textbf{u}$, and $p$ stand for time, velocity, and pressure, respectively. The physical parameters correspond to the liquid density $\rho$, viscosity, $\mu$, surface tension, $\sigma$, surfactant-free surface tension, $\sigma_s$, and  gravitational acceleration, $g$; $T=\sqrt{\rho R_o^3/\sigma_s}$ is the capillary time scale, hence the velocity scale is $U=R_o/T= \sqrt{\sigma_s/(\rho R_o)}$. The interfacial surfactant concentration, $\Gamma$, is scaled on the saturation interfacial concentration, $\Gamma_{\infty}$, whereas the bulk and bulk sub-phase (immediately adjacent to the interface) surfactant concentrations given by $C$ and $C_s$, respectively, are scaled on the initial bulk surfactant concentration, $C_\infty$.
As a result of this scaling, the dimensionless equations read 
\begin{equation}
 \nabla \cdot \tilde{\textbf{u}}=0,
\end{equation}
\begin{equation}
\tilde{\rho} (\frac{\partial \tilde{\textbf{u}}}{\partial \tilde{t}}+\tilde{\textbf{u}} \cdot\nabla \tilde{\textbf{u}}) + \nabla \tilde{p}  =  -Bo \textbf{i}_z + Oh ~ \nabla\cdot  \left [ \tilde{\mu} (\nabla \tilde{\textbf{u}} +\nabla \tilde{\textbf{u}}^T) \right ] +
\int_{\tilde{A}\tilde{(t)}} 
(\tilde{\sigma} \tilde{\kappa} \textbf{n} +   \nabla_s  \tilde{\sigma})  \delta \left(\tilde{\textbf{x}} - \tilde{\textbf{x}}_{_f}  \right)\mbox{d}\tilde{A},
\end{equation}
\begin{equation} 
\frac{\partial \tilde{C}} {\partial \tilde{t}}+\tilde{\textbf{u}}\cdot \nabla \tilde{C}= \frac{1}{Pe_b} \nabla^2 \tilde{C},
\end{equation}
 \begin{equation} 
 \frac{\partial \tilde{\Gamma}}{\partial \tilde{t}}+\nabla_s \cdot (\tilde{\Gamma}\tilde{\textbf{u}}_{\text{t}})=\frac{1}{Pe_s} \nabla^2_s \tilde{\Gamma}+ Bi \left ( k  \tilde{C_s} (1-\tilde{\Gamma})- \tilde{\Gamma}  \right ),
 \end{equation}
\begin{equation} 
\tilde{\sigma}=1 + \beta_s \ln{\left(1 -\tilde{\Gamma}\right)},
\label{marangoni_eq}
 \end{equation}
\noindent
where the density and viscosity are given by $\tilde{\rho}=\rho_g/\rho + \left(1 -\rho_g/\rho\right) \mathcal{H}\left(\tilde{\textbf{x}},\tilde{t}\right)$ and $\tilde{\mu}=\mu_g/\mu+ \left(1 -\mu_g/\mu\right) \mathcal{H}\left( \tilde{\textbf{x}},\tilde{t}\right)$
wherein $\mathcal{H}\left( \tilde{\textbf{x}},\tilde{t}\right)$ represents a smoothed Heaviside function, which is zero in the gas phase and unity in the liquid phase, while the subscript $g$ designates the gas phase;
$\tilde{\textbf{u}}_{\text{t}}= \left ( \tilde{\textbf{u}}_{\text{s}} \cdot \textbf{t} \right ) \textbf{t}$ is the tangential velocity at the interface in which $\tilde{\textbf{u}}_{\text{s}}$ represents the interfacial velocity; $\kappa$ is the curvature; $\nabla_s=\left({\mathbf{I}}-\mathbf{n}\mathbf{n}\right)\cdot \nabla$ is the interfacial gradient wherein $\mathbf{I}$ is the identity tensor and $\mathbf{n}$ is the outward-pointing unit normal; $\delta$ is the Dirac delta function, which is equal to unity at the interface, located at $\tilde{\mathbf{x}}=\tilde{\mathbf{x}}_f$, and zero otherwise; $\tilde{A} (\tilde{t})$ is the time-dependent interface area. 

The dimensionless groups that appear in the above equations are defined as
\begin{equation}
Bo=\frac{\rho g R_o^2}{\sigma_s }, ~~~
La=\frac{1}{Oh^2}= \frac{\rho \sigma_s R_o}{\mu^2 },    
\end{equation}
\begin{equation}\label{surf_param}
Bi=\frac{k_d R_o}{U},        ~~~
k=\frac{k_a C_\infty }{k_d}, ~~~
Pe_s=\frac{ U R_o}{D_s},     ~~~ 
Pe_b=\frac{ U R_o}{D_b},     ~~~ 
\beta_s= \frac{\Re T \Gamma_\infty }{\sigma_s },
\end{equation}
where $Bo$ stands for the Bond number and represents the ratio of gravitational to capillary forces; 
$Oh=\mu/\sqrt{\rho\sigma_s R_o}$ is the Ohnesorge number that measures the relative importance of viscous to surface tension forces, and $La$ is the Laplace number; $Bi$ denotes the Biot number representing the ratio of characteristic desorptive to convective time-scales; $k$ is the ratio of adsorption to desorption time scales; $Pe_s$ and $Pe_b$ are the interfacial and bulk Peclet numbers, respectively, and represent the ratio of convective to diffusive time-scales in the plane of the interface and the bulk, respectively; and $\beta_s$ is the surfactant elasticity number, which measures the sensitivity of the surface tension to the surfactant concentration. The chosen density and viscosity ratios,  $\rho_g/ \rho =1.2 \times 10^{-3}$ and $\mu_g/\mu= 0.018$, respectively, are representative of an air-water system. 

The Marangoni stress, $\tilde{\tau}$, is expressed as a function of $\tilde{\Gamma}$:
 \begin{equation}
 \label{langmuir}
 \tilde{\tau} \equiv  \nabla_s \tilde{\sigma} \cdot  \textbf{t} =-\frac{\beta_s}{1 -\tilde{\Gamma}} \nabla_s\tilde{\Gamma} \cdot \textbf{t},
\end{equation}
where $\mathbf{t}$ is the unit tangent to the interface; tildes are dropped henceforth.
The Marangoni time-scale, $\mu R_o/ \Delta \sigma =O(10^{-2})$ s, as compared to the  capillary and sorptive time-scales, which are of order  $1$ s;  thus Marangoni stresses will play a key role in the dynamics. 

The dimensionless computational domain size is chosen as $(15R_o)^3$, which is found to be sufficiently large to render boundary effects negligible. Hence a radial component is defined as $r=\sqrt{\left(x-x_o\right)^2 + \left(y-y_o\right)^2}$ where $x_o$ and $y_o$ are the abscissa and ordinate bubble position, respectively. Solutions are sought subject to Neumann boundary conditions on all variables at the lateral boundaries, $p=0$ at the top boundary $z=15R_o$, and no-slip at  $z=0$. At the free surface, we impose $\textbf{n}\cdot\nabla \tilde{C}=-BiPe_b \left ( k  \tilde{C_s} (1-\tilde{\Gamma})- \tilde{\Gamma}  \right )$ as a condition on $\tilde{C}$. Simulations are initialised as a bubble resting immediately beneath the free surface before its rupture. Its initial shape is determined by solving the Young-Laplace equation for  $Bo=10^{-3}$, which is sufficiently small to avoid the effect of gravity on the bubble bursting process. To initialise bursting, the top spherical cap of the bubble is removed, leaving only a bubble cavity (see figure \ref{configuration}a). A similar approach has been used by \cite{Boulton-Stone_jfm_1993}, \cite{Garcia-Briones_ces_1994}, \cite{Duchemin_pof_2002} and \cite{Deike_prf_2018}. Our numerical simulations have been validated against the work of \cite{Eggers_prl_1993} and \cite{Lai_prl_2018} in terms of liquid thread breakup, and the scaling of the ejected jet length, $L_d$, with $La$ (see figure \ref{configuration}b), respectively.

\section{Results\label{sec:Results}}
\begin{figure}
\begin{center}
\includegraphics[width=1.1\linewidth]{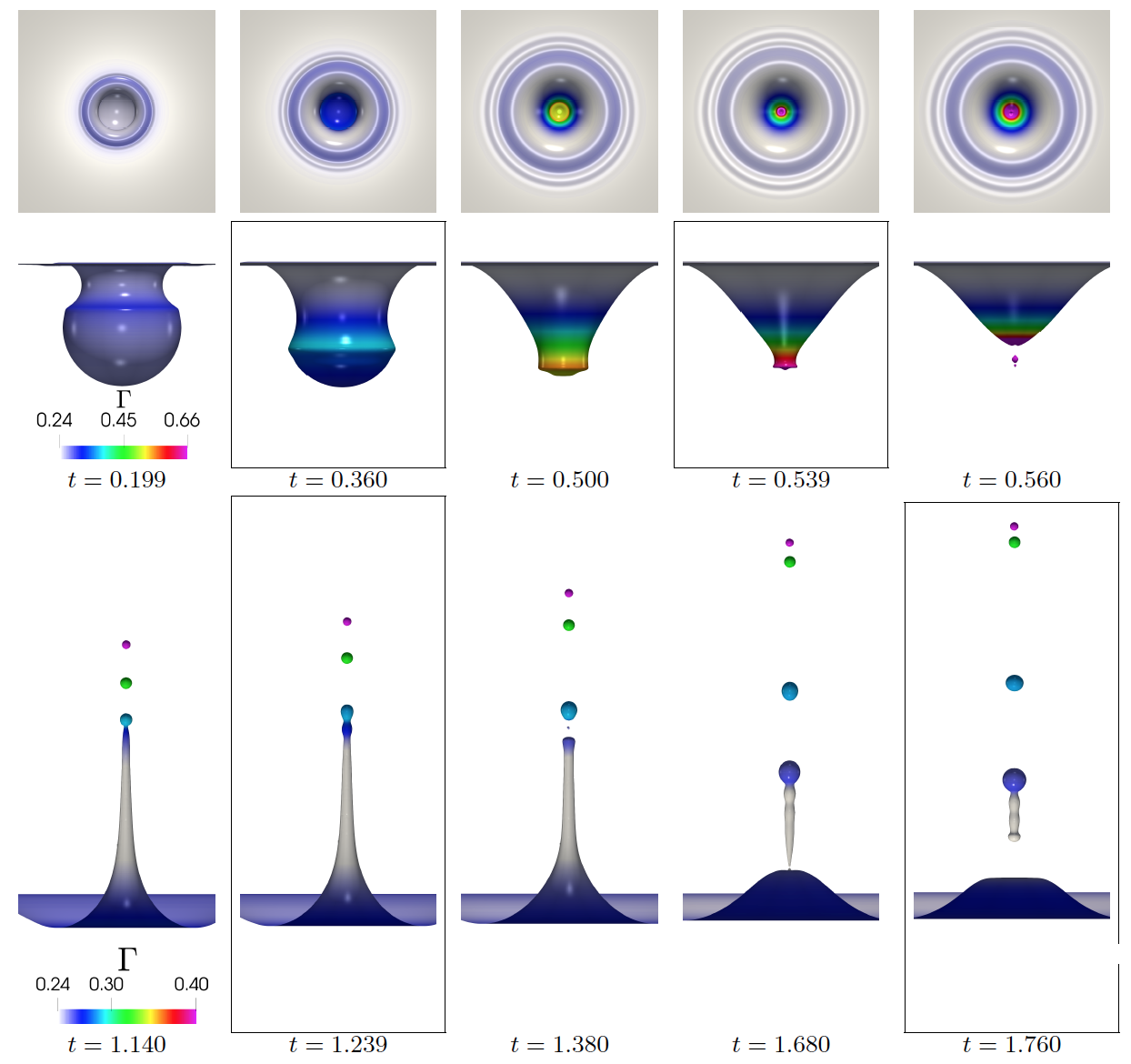}
\end{center}
\caption{\label{3d_temporal} Spatio-temporal evolution of the dynamics of the interface and of the interfacial   surfactant concentration, $\Gamma$, with $La=2\times10^4$, $Bo=10^{-3}$, $Pe_s=1$, $\beta_s =0.9$, $Bi=0.01$, $k=1$, and
$\Gamma_o =k/4$. The colour bars indicate the magnitude of $\Gamma$. Top row: top-view of the interface; middle row: side-view of cavity collapse; bottom row: Worthington jet (entrapped bubble is not shown). }\end{figure}

\begin{figure}
\includegraphics[width=1\linewidth]{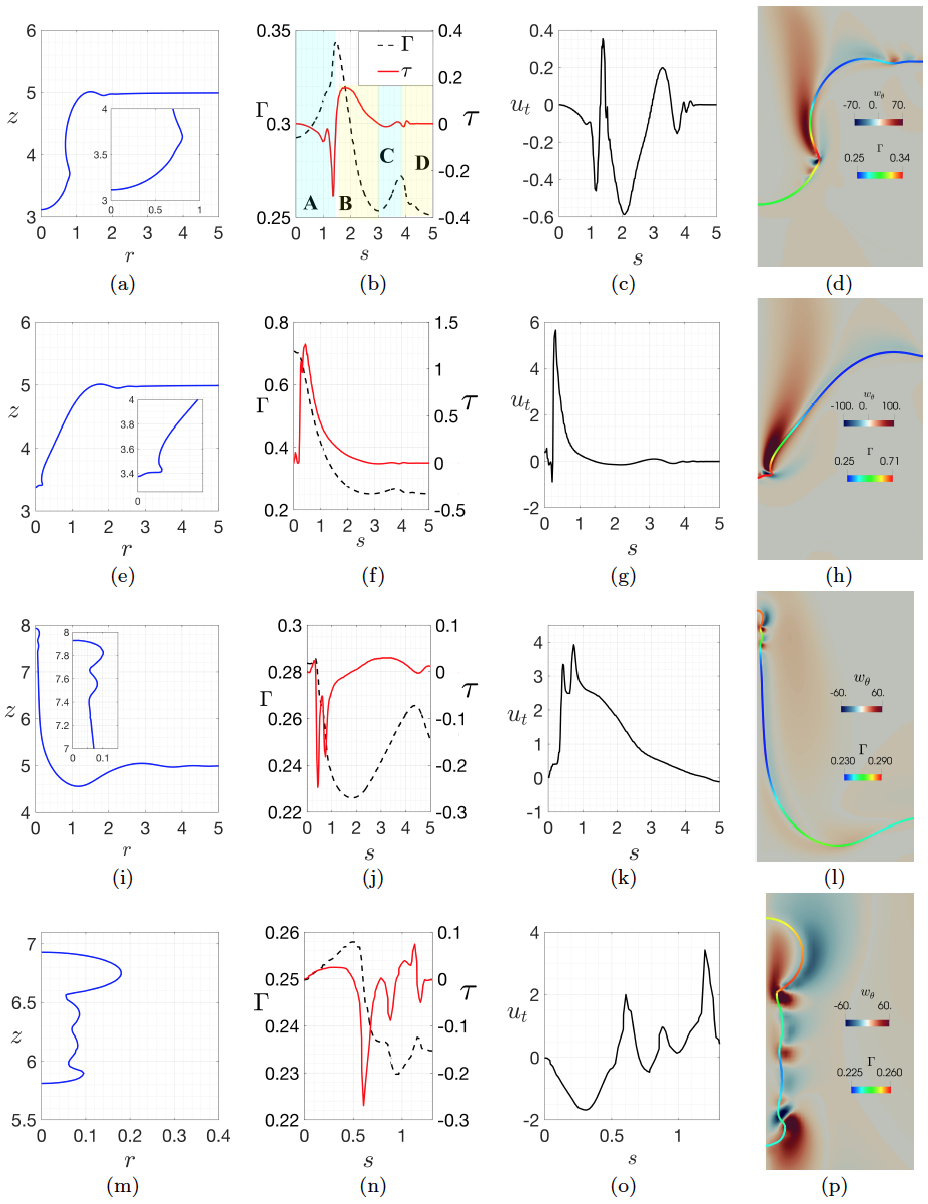}
\caption{\label{marangoni} Interface location, $\Gamma$ and $\tau$, $u_t$, and $\omega_{\theta}$ are shown in columns one to four, respectively. In columns 1,4 and 2,3, the variation is with respect to the dimensionless radial coordinate, $r$, and arc length, $s$, respectively. Panels (a)-(d), and (e)-(h) show the cavity collapse dynamics $t=0.360$ and $t=0.539$, respectively, (i)-(l) the Worthington jet at $t=1.239$, and (m)-(h) the retracting ligament at $t=1.760$. The insets in (e) and (i) respectively focus on the bottom of the collapsing cavity, and the oscillations at tip of the Worthington jet that will eventually lead to its breakup. A description of regions `A'-`D' in (b) is provided in the text. The parameter values are the same as in Fig. \ref{3d_temporal}.
}
\end{figure}

The spatio-temporal evolution of the interfacial dynamics, shown in figure \ref{3d_temporal}, is considered for the case characterised by $La=2\times10^4$, $Bo=10^{-3}$, $Pe_s=1$, $\beta_s =0.9$, $Bi=0.01$, $k=1$, and $\Gamma_o =k/4$. At early times, a large capillary pressure is generated near the nucleated hole due to the high curvature of the interface joining the spherical bubble and the horizontal free surface. This capillary pressure leads to rapid expansion of the hole and the creation of a toroidal capillary wave, which travels toward the bottom of the bubble. 
The convergence of this wave on the bubble rear leads to the formation of a cusp-like region, which is relieved via the 
detachment of downward-moving conical bubbles, and the formation of a vertical, upward-directed, high-speed, Worthington jet.  Small droplets are subsequently ejected from the tip of the jet, triggered by a Rayleigh-Plateau instability. A pinchoff event is also seen to occur at the base of the Worthington jet that leads to retraction of the emitted ligament into a spheroidally-shaped drop. 

In figure \ref{marangoni}, we show snapshots of the interface, the interfacial concentration, $\Gamma$, the Marangoni stress, $\tau$, the interfacial tangent velocity component, $u_t$, that provides a measure of mobility, and the azimuthal component of the vorticity, $\omega_\theta$. These snapshots, which are taken at $t=0.360$, $0.539$, $1.239$, and $1.760$, corresponding to the framed panels in figure \ref{3d_temporal}, reflect the strong coupling between these flow variables. Figure \ref{marangoni}a,b illustrates that the collapsing cavity, and the accompanying capillary wave, transport surfactant towards the bottom of the bubble giving rise to a local decrease in surface tension. It is also instructive to separate figure \ref{marangoni}b into four regions due to the existence of stagnation points. In region `A', $\tau < 0$ and $u_t < 0$, indicating that the direction of the Marangoni flow is towards the bottom of the bubble, which aids cavity collapse and surfactant transport in this direction. In region `B', $\tau >0$ and $u_t > 0$, thus the Marangoni flow is directed away from the origin, which retards the collapse process. A similar behaviour is seen in regions `C' and `D' in which the trailing edge of the capillary wave has a similar $\Gamma$ distribution, albeit with a smaller magnitude, to its leading edge. As a result, $\tau < 0$ and $\tau > 0$, and $u_t < 0$ and $u_t > 0$ in regions `C' and `D', which drives flow towards the bottom of the bubble and its tail, respectively. Figure \ref{marangoni}b,c also shows the existence of an interfacial stagnation point ($s \sim 1.4 $), where surfactant accumulates, $\Gamma$ is highest, and the magnitude of $\tau$ is largest. This occurs in the region where $\omega_{\theta}$ changes sign as the stagnation point is created (see figure \ref{marangoni}d) corresponding to the capillary wave moving towards the bottom of the bubble.

Figure \ref{marangoni}e-h shows the dynamics at $t=0.539$  prior to the cavity collapse. The stagnation point where the surfactant accumulates has moved downward to the bottom of the cavity and the magnitude of $\omega_{\theta}$ has increased in comparison to that in figure \ref{marangoni}d. The interfacial surfactant concentration reaches its maximum value as the surfactant-laden capillary wave converges on the flow origin. The Marangoni stresses drive motion from high to low $\Gamma$ regions and therefore act to oppose the flow, as indicated by the fact that both $\tau >0$ and $u_t>0$ over the majority of the spatial domain except in the close vicinity of the bottom of the cavity.  

Figure \ref{marangoni}i shows a snapshot of the jet for $t=1.239$ in a situation of pinch-off `escape' with a bulbous region formed at its tip. From figure \ref{marangoni}j, we observe that $\Gamma$ has two peaks: one at the jet tip, and a lower one far upstream. Figure \ref{marangoni}l also shows that the change in the sign of $\omega_{\theta}$ is linked to the $\Gamma$ distribution in the bulbous region. The associated $\tau$ distribution is such that $\tau <0$ between the bottom of the jet and close to the bulbous region at the jet tip (that is for $0.4 <s< 1.9$), and $\tau >0$ elsewhere including in the bulbous region. Flow is driven by capillarity from this region towards the quasi-cylindrical jet body and, from figure \ref{marangoni}k, it is seen that $u_t > 0$ over the whole domain. This suggests that Marangoni stresses oppose this capillary-induced motion driving flow from the bottom of the jet towards the bulbous region. When these stresses are not sufficiently strong to overcome the Rayleigh-Plateau instability, an ejection of a droplet occurs after the pinch-off of the jet tip. The first drops detached from the tip are characterised by high interfacial concentration while successive droplet detachments have lower $\Gamma$, as demonstrated in figure \ref{3d_temporal} for $t\ge1.114$.   

At $t=1.680$ (see figure \ref{3d_temporal}), the Worthington jet pinches off from its base forming an elongated ligament thread. Capillary waves develop on the ligament surface, leading to interfacial oscillations as the detached ligament transitions to a spherical drop. In figure \ref{marangoni}m-p, we have isolated the retracting ligament shown at $t=1.760$ from the rest of the flow. In figure \ref{marangoni}o, we observe the formation of four stagnation points at $s=0$ 0.55, 0.7, and 0.84, which are connected to the change in the distributions of $\Gamma$ and $\tau$. From figure \ref{marangoni}n,o, it is also seen that the Marangoni stresses oppose ligament pinchoff since $\tau > 0$ and $u_t < 0$ ($\tau <0$ and $u_t > 0$) for $0 \leq s \leq 0.55$ ($0.55 \leq s \leq 0.7$). For $0.7 \leq s \leq 0.84$ ($s > 0.84$), the Marangoni stresses oppose (aid) the stretching of the ligament as $\tau <0$ and $u_t >0$ ($\tau >0$ and $u_t >0$). Close inspection of $\omega_{\theta}$ (see figure \ref{marangoni}p) reveals that  high vorticity production is observed close to the first stagnation point, which corresponds to the ligament `neck'. As shown by \cite{Constante-Amores_prf_2020}, the presence of such high vorticiy regions near the neck is a requirement for the Marangoni-driven inhibition of capillary-induced `end-pinching' of retracting surfactant-laden ligaments.


\begin{figure}
\begin{center}
\includegraphics[width=1.05\linewidth]{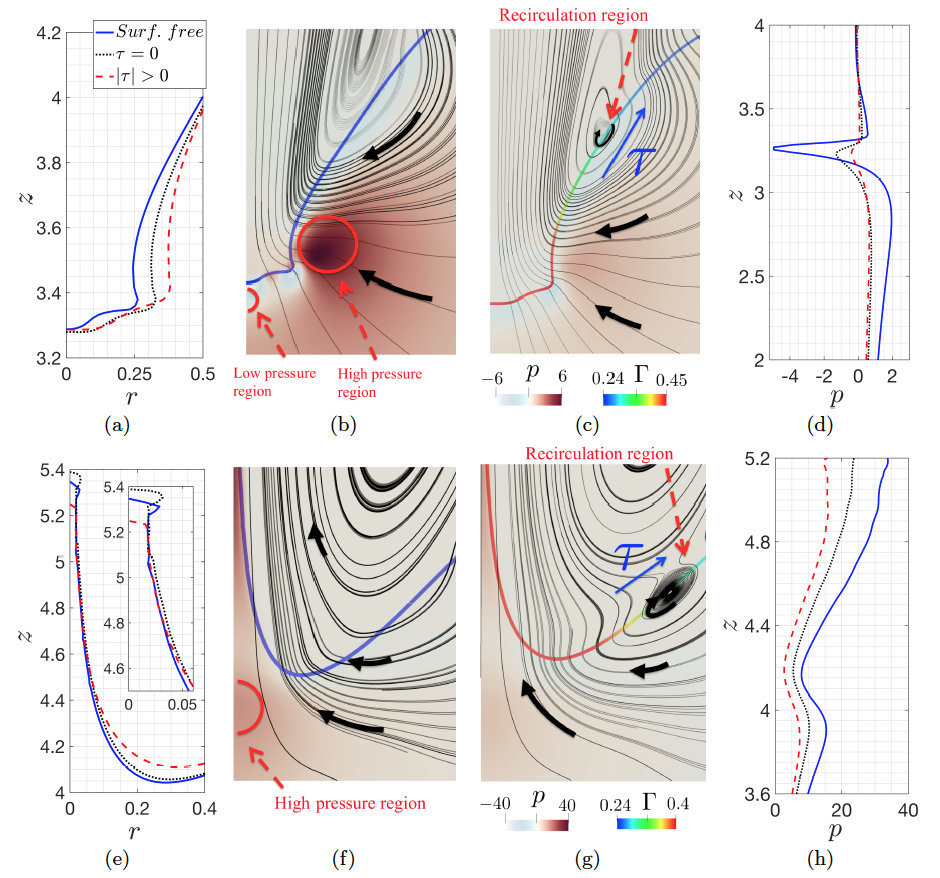}
\end{center}
\caption{\label{fig:Marangoni} Interface location, pressure field together with a representation of the streamlines for the surfactant-free and full-Marangoni ($|\tau|>0$) 
cases, and pressure on the axis of symmetry, are shown in columns one to four, respectively. Panels (a-d) and (e-h) show the cavity collapse dynamics (at $t=0.440$, 0.496, and 0.500, for the surfactant-free,  no-Marangoni [$\tau =0 $], and full-Marangoni cases) and the Worthington jet (at $t=0.507$, 0.594, and 0.619, for the surfactant-free, $\tau =0 $, $|\tau| >0 $ cases), respectively. The parameter values are the same as those used to generate figure 2.}
\end{figure}

In order to isolate the surface tension-reducing effects of surfactants from those associated with Marangoni stress-formation, we
show in figure \ref{fig:Marangoni}a snapshots of the interface at $t=0.44$, 0.496, and 0.500 for the surfactant-free, surfactant-laden but Marangoni-suppressed, and full-Marangoni cases, respectively, prior to cavity-collapse for the same parameters as in figure \ref{marangoni}. These times were chosen at  identical spatial locations of the bubble rear in the axis of symmetry $(z \sim 3.28)$. For the Marangoni-suppressed case, the reduced surface tension value is calculated via replacing $\Gamma$ by $\Gamma_o$ in equation \ref{marangoni_eq}. The presence of surfactant has been shown to enhance capillary wave-damping by \cite{Asaki_prl_1995} due to the interfacial rigidification brought about by $\tau$, and this is seen clearly in figure \ref{fig:Marangoni}a: the size of the cavity is largest for the full-Marangoni case, at the same stage of the dynamics for the three cases considered. 

The capillary pressure field shown in figure \ref{fig:Marangoni}b,c for the surfactant-free and full-Marangoni cases is influenced heavily by the capillary pressure and, therefore, the local interfacial curvature and surface tension. From panels (b) and (c) of figure \ref{fig:Marangoni}, it is seen  that the pressure is highest immediately upstream of the capillary wave peak and this pressure gradient drives flow towards the lower-pressure region located at bottom of the cavity that coincides with the axis of symmetry. Furthermore, Marangoni stresses induce a recirculation zone close to the free surface as shown in figure \ref{fig:Marangoni}c. In figure \ref{fig:Marangoni}d, we also show the axial distribution of the pressure at the axis of symmetry and this displays a peak at the interface due to capillarity. Owing to the presence of surfactant, the surface tension is reduced, which leads to a concomitant fall in the pressure, as illustrated via comparison of figure \ref{fig:Marangoni}b,c. The pronounced reduction in capillary pressure, due primarily to the accumulation of surfactant at the bottom of the cavity, is shown in figure \ref{fig:Marangoni}d. 

As mentioned above, the convergence of the capillary wave on the bottom of the cavity leads to the formation of a  Worthington jet \citep{Gordillo_jfm_2019}, which is shown in figure \ref{fig:Marangoni}e for the surfactant-free, Marangoni-suppressed, and full-Marangoni cases for $t=0.507$, $t=5.94$, and $t=0.619$, respectively; again, these times are chosen  at  nearly identical spatial locations of the jet tip in the axis of symmetry. The larger pressure gradient associated with the surfactant-free case, discussed above, leads to longer jets, with more pronounced bulbous regions at their tips, in comparison to the full-Marangoni case where the retarding Marangoni stresses induce a recirculation zone close to the jet-base. Close inspection of figure 4e also reveals that the longest jets are associated with the $\tau =0$ (rather than the surfactant-free) case wherein there is no Marangoni-induced retardation and for which $La$ is lowest. As shown previously \citep{Lai_prl_2018}, decreasing $La$ leads to faster, and thinner jets, with a greater propensity for breakup. As a result, a reduction in the number of ejected droplets is observed for the shorter, and slower, full-Marangoni jets: four (see figure 2), seven, and nine droplets (the latter two not shown) for the full-Marangoni, surfactant-free, and Marangoni-suppressed cases, respectively. Finally, as shown in figure \ref{fig:Marangoni}h, there is an adverse pressure gradient in all cases, since capillarity drives flow from the bulbous region towards the jet base, which is largest for the surfactant-free case. 

\begin{figure}
\begin{center}
\includegraphics[width=1\linewidth]{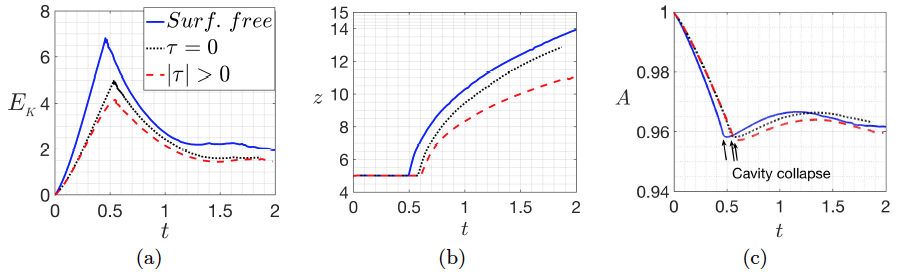}
\end{center}
\caption{\label{kinetic} Temporal evolution of kinetic energy, the maximal vertical interface location, and interfacial area (normalised by its initial value),  (a)-(c), respectively, for the surfactant-free, full-Marangoni, and no-Marangoni cases, for the same parameters as in figure \ref{3d_temporal}.
}
\end{figure}

The immobilising effect of the Marangoni stresses can be seen in figure \ref{kinetic}a in which we plot the kinetic energy, defined as $E_k=\int_V (\rho \textbf{u}^2/2) dV$, where Marangoni stresses reduce the maximal and asymptotic values of $E_k$ in comparison to the surfactant-free and no-Marangoni cases. In figure \ref{kinetic}b, we  observe further that the motion of the interface is retarded maximally when Marangoni stresses are enabled fully. Inspection of figure \ref{kinetic}c also reveals that the interfacial area $A$, normalised by its initial value, $A_o$, reduces over time leading to large $\Gamma$ at the moment of interfacial vertical collapse at $t \sim 0.550$. 

\section{Concluding remarks}\label{sec:Con}
 
The effect of Marangoni-induced flow, brought about by the presence of surfactant, on the dynamics of  a bubble bursting through an interface was studied using a hybrid front-tracking/level-set method. Our results indicate that a surfactant-covered toroidal  capillary wave forms, following the collapse of the cavity, whose motion is retarded by the surfactant-induced Marangoni stresses; these stresses drive flow from regions of high surfactant concentration (low surface tension) to low concentration (high tension) regions. The immobilising effect of the surfactants due to the Marangoni stresses is also observed via the marked reduction in the system kinetic energy and the  generation of shorter, and slower, Worthington jets. The breakup of these jets is accompanied by the formation of fewer droplets in comparison to the surfactant-free case.

\subsection*{Acknowledgements}

This work is supported by the Engineering \& Physical Sciences Research Council (EPSRC), United Kingdom, through a studentship in the Centre for Doctoral Training on Theory and Simulation of Materials at Imperial College London funded by the EPSRC (EP/L015579/1) (Award reference:1808927), and through the EPSRC MEMPHIS (EP/K003976/1) and PREMIERE (EP/T000414/1) Programme Grants. We also acknowledge HPC facilities provided by the Research Computing Service (RCS) of Imperial College London for the computing time. The numerical simulations were performed with code BLUE \citep{Shin_jmst_2017} and the visualisations have been generated using ParaView.

\bibliographystyle{jfm}

\end{document}